# Autonomous Weapons Systems

Spotlight Session 4

As we have heard throughout the conference so far, we are witnessing bitter and brutal armed conflicts in many places around the world, and facing extremely challenging humanitarian situations. People need help right now: protection, food, water, shelter, medical treatment and more. So I want to begin our discussion by asking each of our panelists: why are we talking about autonomous weapon systems? Why is it important, given everything else that is going on? Why should governments and national Red Cross and Red Crescent societies be devoting time and attention to this issue?

We are talking about autonomous weapons systems because robotics, autonomous systems data analytics and artificial intelligence are changing the way wars are being fought and potentially how they are determined, how they are evaluated and how consequences for their use is metered out. We see AI-enabled commercial off the shelf drones being adapted and used in conflicts across the world from Ukraine, to Syria, to Yemen, Sudan and Myanmar. We see digital infrastructure being invested in by wealthy militaries including vast data networks, surveillance, and decision support tools. We see great powers developing advanced weapons systems capable of leveraging digital infrastructure to improve decision agility. We see middle-powers working with allies to adapt limited capabilities to deliver greater lethality and scale of effects. Governments, the national Red Cross and Red Crescent societies should be devoting time and attention to this issue because autonomous weapons systems change the way humans make decisions, the effect of those decisions and who is accountable for decisions made. We must remain vigilant, informed and human-centred as we tackle our deliberations on developing norms regarding their development, use and justification.

## Autonomous weapons systems infrastructure and supply chain operational dependencies and ethical obligations.

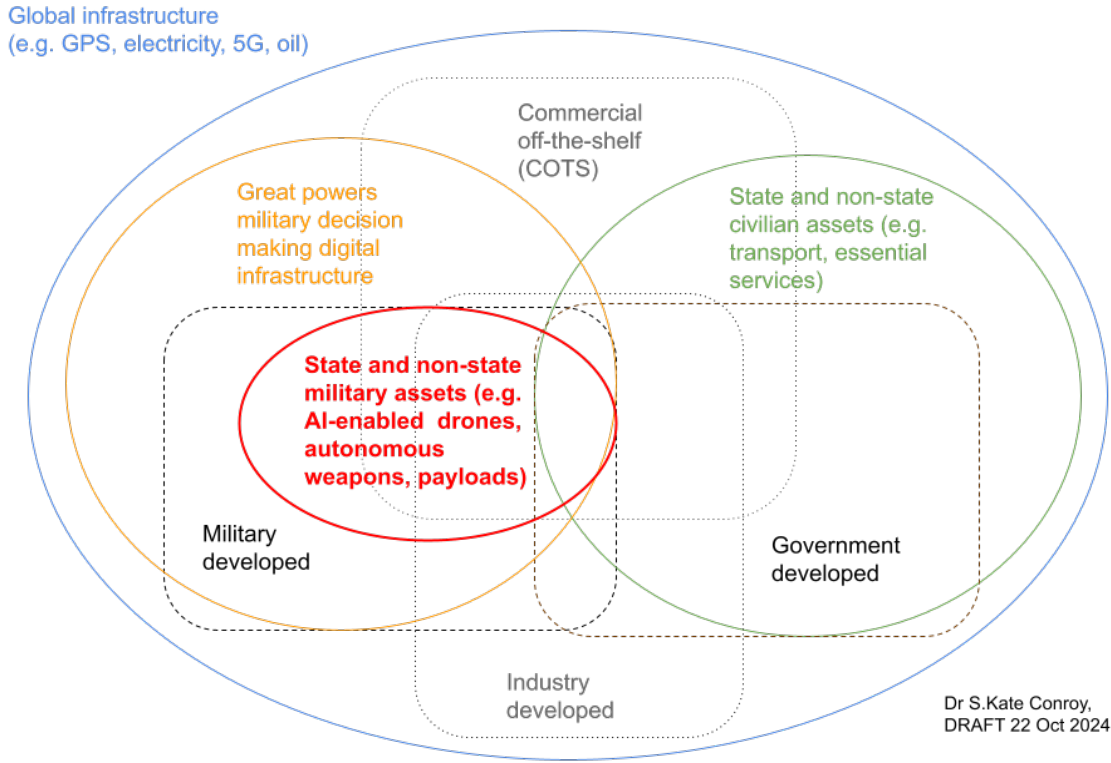

Dr S.Kate Conroy, DRAFT 22 Oct 2024

Autonomous weapons systems are deployed with dependencies on global, military and civilian infrastructure and built using a mix of military, government and industrial supply chains. Wealthy militaries have more control over infrastructure and supply chains than less well off militaries. The ethical obligations on wealthy militaries extend to ensuring responsible development and use of the enabling infrastructure of AI-enabled and autonomous weapons systems. Poorer state and non-state militaries will depend on the decision infrastructure they have access to either from global commercial or government-controlled assets and supply chains or via allies. The ethical obligations on poorer militaries likely to be at a tactical rather than a strategic and infrastructural decision making level.

In the technical briefing, we heard that some people argue that autonomous weapons have the potential to actually be better at complying with IHL than humans and will lead to better humanitarian outcomes – that they will be more precise than existing weapons, and will be able to scan and assess a situation much more quickly and accurately, making a better decision about the legality of an attack. Kate, can I ask you, from a technical point of view, how plausible is this?

Thank you Anne. I agree that autonomous weapons have the potential to be better at complying with IHL and could lead to better humanitarian outcomes. From a technical point of view, all of this is possible, should human beings invest in these capabilities to prioritise human-centred values in the way systems are designed and used. My concern lies not with the technical possibilities but the lack of investment and interest in control mechanisms that would make a genuine impact on the conduct of war.

In my comments on this question, I would like to achieve two outcomes, provide some thoughts on how to improve autonomous weapons compliance with IHL and then discuss the plausibility (including barriers) of implementing these suggestions.

**Ways to improve autonomous weapons compliance in IHL include:**
- Upskill weapons decision makers in IHL and what is obliged when developing and deploying any weapons system to ensure legal compliance.
- Developing best practice in weapons reviews including requirements for industry to ensure that any new weapon, means or method of warfare is capable in being used lawfully.
- Develop human-centred test and evaluation metrics, tests and scenarios that incorporate IHL relevant factors to enable adequate evaluation of weapons systems pre-deployment and AI impact assessments in the intended context of operations.
- Invest in digital infrastructure to increase knowledge and situational awareness of the civilian environment in a conflict and its dynamics.
- Invest in research on the real effects and consequences of civilian harms to the achievement of military and political objectives and the potential for peace, prosperity and security.
- Improve secure communications between stakeholders in a conflict (including civilians, NGOs, governments and militaries) to improve the fidelity of civilian environment models and military decision-making
- Upskill governments and national Red Cross and Red Crescent societies in what is technically achievable with emerging technologies so that they can contribute to system requirements, test and evaluation protocols and operational rules of use and engagement.

**Possibility**

Now, given that all of these suggestions are plausible, what of these suggestions is likely or easy to achieve, what is of medium difficulty and what will be difficult and expensive?

*Easy*
What is easiest to achieve is reinforcing existing mechanisms for compliance with IHL including Article 36 weapons reviews. Many of the ethical concerns of autonomous weapons systems can be absolved if weapons are developed in a way that ensures they can be used lawfully.

*Medium*

What is harder, but still achievable is ensuring operators use autonomous systems in operational contexts with a human-centric, ethical mindset. Increasing human understanding of the impacts of autonomous systems on the lives of those in a conflict is imperative to encourage their use in an ethically-minded as well as lawful ways. Let us not forget that weapons can be deployed lawfully that nevertheless deliver effects that are shocking and horrifying to those who experience them. Autonomy is likely to change the way force is deployed and will always need to be monitored, questioned and adapted.

*Difficult*

What is difficult, but we must continue to require it, is significant investment in the methods to improve civilian protection that goes far beyond what is legally obliged to delivering to our highest ethical expectations for the conduct in war. New technologies allow an unprecedented ability for militaries to control their use of force to protect civilians and protected objects, but to achieve this ambition they must seriously invest in human-centred interfaces, high fidelity digital decision environments updated in real time to ensure human decision makers fully understand the civilian environment within which weapons systems are deployed, to test the ethicality of decisions including what the consequences are likely to be by the use of force within this environment including humanitarian, political, economic, and societal consequences.

> As well as concerns about IHL and the protection of civilians, there are also ethical questions about the use of autonomous weapons. I personally find the idea of giving machines the power to make life or death decisions utterly horrifying. But then war is already utterly horrifying – will autonomous weapons really change that, ethically speaking? Kate, could you take us through the ethical aspects of this?

Thanks Anne, I appreciate that you find the idea of giving machines the power to make life or death decisions utterly horrifying, and you acknowledge that war is already utterly horrifying. So, how will autonomous weapons change the nature of war and can they possibly make it less horrifying? If it makes it more horrifying, in what ways does it enable such?

The horror of autonomous systems comes from three axes, 1) *The Machine Mind*: Machines having their own ideas of who or what to harm, 2) *The Missing Human*: humans not having a role in these decisions and 3) *The Inadequate Machine*: Machines not knowing enough to be accurate in their decisions.

**The Machine Mind**

One of the most frightening ideas of autonomous weapons systems is the idea of machines having a mind of their own and being unleashed amongst human beings; authorised and empowered to make decisions to harm them. International humanitarian law, while not forbidding autonomous functions in weapons systems, does require an appropriate relationship to maintain between human command and the behaviour of a weapons system such that humans are always responsible for their use of these systems. While science fiction stories focus on the worse possibilities. It is very important to reinforce and support adherence to IHL with the use of any new weapon, means or method of warfare. Responsible militaries already have these mechanisms in place. In reality, no organisation wants uncontrollable autonomous systems that they do not understand deployed on the battlefield. If we watch the way that robotics, autonomous systems and AI are being used today and likely to be used tomorrow 'machine minds' are all designed, developed, deployed and controlled by humans. The horror of these systems can occur when their operations are well known, they are controlled by humans and delivering exactly what was intended by human operators.

**The Missing Human**

Perhaps, some might say, alright, humans have designed these systems, but are they are abdicating their role in decisions to use force when using them? Again, IHL helps us a lot because it requires not only that human commanders take responsibility for the use of a weapons system, they must also know enough about how it works and why it works to be morally and legally responsible for its deployment. The use of most modern weapons systems involves a spatiotemporal gap between the time of a decision and the impact of that decision. Adding autonomous functions into that spatiotemporal gap does require ensuring human beings understand the parameters of those functions and can vouch for the behaviour of the system when its operators outside of the envelop of intervention by human beings. However, ensuring this is all under human control.

**The Inadequate Machine**

Alright, you say, so perhaps machines are not making up their own minds and humans are in control, but still, these autonomous weapons systems are making decisions badly. They are using patchy data to extrapolate on what is a lawful target and what isn't and we don't want unreliable weapons making targeting decisions. There is no doubt that machine learning-fed systems can be very unreliable and humans must not get caught in the trap of believing the marketing and hype of defence industries about the precision and accuracy of these systems simply because mystical 'AI' is involved. However, autonomous weapons systems can be made highly reliable in a narrow range of lawful behaviours (e.g. see the Closed in Weapons System, CWIS) and highly reliable when integrated with great care with human beings and other systems to increase human awareness and understanding of the circumstances of a decision. The epistemic issues with autonomous weapons systems are a matter of how they are developed, how they are tested and evaluated and how well they are informed, updated and limited in deployment. The inadequacy of the machine is the inadequacy of human beings to ensure autonomous systems behave as an extension of human intent informed by our values as a society.

At the end of the day, some of the most horrifying aspects of autonomous weapons systems is that they seem like monsters, unknowable unethical agents capable of terrible means and ends. Yet, they are not their own monsters, they are extensions of human will, manifestations of human values and deliberative decision making. If they are monstrous, then it is us who are the monsters and we must be elevated, transformed and constrained.

## Kate, from an academic perspective, from the perspective of those actually developing, evaluating and deploying AI and automated systems, what factors are important to consider in evaluating mechanisms for the regulation of autonomous weapons? What is feasible for governments and the UN to take ownership of and be trusted with? Is it even possible?

When evaluating mechanisms for the regulation of autonomous weapons, there are key factors to take seriously:

(1) *Low regulated enablers*: We need to understand the role of largely unregulated corporations in enabling the use of autonomous weapons both ethically and problematically.
    a. Using autonomous weapons reliably requires access to data and communications usually enabled by private companies who serve civilian customers as well militaries such as Palantir, Google, Microsoft, SpaceX (Starlink), and Amazon.
    b. These corporations work hard to ensure that they can continue to do business and there are usually carve outs to allow them to work within militaries differently to the way they supply services to governments,
        i. E.g. the obligations of the EU AI Act do not apply to military use
    c. Private companies create commercial off the shelf robotics technologies such as drones that are increasingly used in conflicts as an affordable way for militaries (both state and non-state) to use autonomous aspects in the delivery of effects.
(2) *Diverse systems*: Autonomous weapons systems are not one physical or chemical thing or one type of mechanism or function. Autonomy can be used to refer to different sorts of metaphysically bounded entities with many-to-many computational, algorithmic and material instantiations teamed with human beings and deployed in different circumstances.
(3) *Interpretable requirements*: The requirements asked of autonomous weapons systems are interpretable. E.g. what does 'predictability' require? It's not as simple as you think. What is predictable enough is likely subjective both by decision makers and organisations depending on military objectives. Focusing on predictability is unlikely to reduce awful outcomes. Arguably some of the most shocking harms to civilians enabled by AI in current conflicts are meticulously planned and executed by humans in the loop—the AI was limited and did exactly as asked. The harms were 100% designed and enabled by humans using technologies in the way they determined was necessarily to achieve their objectives.
(4) *Contextually sensitive ethics*: Autonomous weapons systems will be used very differently and have very different ethical challenges when used by richer or poorer organisations.

(5) *Rapidly changing technologies* Autonomous weapons systems are rapidly developing and adapting.

If we take these five different aspects: low regulated enablers, diverse systems, interpretable requirements, contextual ethics and rapidly changing technologies we see that a regulatory treatment faces significant hurdles. If we add to this the fact that autonomy can be used to improve humanitarian outcomes (if humans choose to do so) as well as the fact that militaries generally do not want to have their decision options unreasonably limited, it makes it additionally challenging.

So, what to do… Governments are responsible for setting requirements for weapons systems. They are responsible for driving ethicality as well as lethality. Governments can require systems to be made and used to better protect civilians and protected objects. The UN can advocate for compliance with IHL, human rights, human-centred use of weapons systems and improved mechanisms to monitor and trace military decision making including those decisions affected by autonomous functionality. The global community should continue to work towards shared best practice and through mechanisms such as the Responsible AI in the Military Domain Call to Action (2023) and Blueprint to Action (2024) and Political Declaration for the Responsible Use of Military AI and Autonomy. We must remain vigilant.

## Contact

Dr S. Kate Conroy née Devitt
E: skateconroy@gmail.com
L: https://www.linkedin.com/in/skateconroy